\documentclass{mic2017}

\usepackage{mathptmx}       
\usepackage{helvet}         
\usepackage{courier}        
\usepackage{type1cm}        
\usepackage{makeidx}         
\usepackage{graphicx}        
\usepackage{multicol}        
\usepackage[bottom]{footmisc}

\makeindex             

\usepackage{graphicx}

\usepackage{amssymb}
\usepackage[misc]{ifsym} 

\usepackage{tabularx} 

\usepackage{verbatim}
\usepackage{array}
\usepackage{color}

\makeatletter
\newif\if@restonecol
\makeatother

\usepackage[ruled,slide]{algorithm2e} 

\usepackage{psfrag}    

\usepackage{subfigure} 

\usepackage{stfloats}  
\usepackage{eqparbox}

\usepackage{color} 

\usepackage{url}  
\usepackage{ifthen}  
\usepackage{multicol}   
\usepackage[utf8]{inputenc} 
\usepackage{mathtools}   %
\usepackage{arydshln}
\usepackage{bm,amsmath}

\newcommand{\mat}[1]{\bm{#1}}

\usepackage{tikz}
\usetikzlibrary{matrix,calc}

\usepackage{amsmath}
\usepackage{eqnarray}

\newtheorem{defn}{Definition}[section]

\begin{document}

\title{On the minimum quartet tree cost problem}
\author{Sergio Consoli\inst{1} \and Jan Korst\inst{1} \and Gijs Geleijnse\inst{1} \and Steffen Pauws\inst{1,2}}
\institute{
 Philips Research, High Tech Campus 34, Eindhoven 5656 AE, The Netherlands,
  \email{\textit{[name.surname]@philips.com}}
   \and
   TiCC, Tilburg University, Warandelaan 2, Tilburg 5037 AB, The Netherlands,
   \email{\textit{s.c.pauws@tilburguniversity.edu}}
}

\id{id}
\maketitle

\begin{abstract}
Given a set of $n$ data objects and their pairwise dissimilarities, the goal of the minimum quartet tree cost (MQTC) problem is to construct an optimal tree from the total number of possible combinations of quartet topologies on $n$, where optimality means that the sum of the dissimilarities of the embedded (or consistent) quartet topologies is minimal. We provide details and formulation of this novel challenging problem, and the preliminaries of an exact algorithm under current development which may be useful to improve the MQTC heuristics to date  into more efficient hybrid approaches.

\end{abstract}

\section{Introduction and problem formulation}

Given a set $N$ of $n\geq 4$ objects, the minimum quartet tree cost
problem (MQTC) deals with a \emph{full unrooted binary tree} with $n$
leaves, a special topology dendrogram having all internal nodes 
connected exactly with three other nodes, the $n$ objects assigned as leaf nodes, and without any distinction between
parent and child nodes~\cite{Furnas84}. A full unrooted binary tree
with $n\geq 4$ leaves has exactly $n-2$ internal nodes, and
consequently it has a total of $2n-2$ nodes. 
Full unrooted binary trees are of
primary interest in clustering contexts because,
of all tree diagrams 
with a fixed number of nodes, they have the richest internal
structure (most differentiated paths between nodes). They are
therefore very suitable 
for representing the structure of a set of objects~\cite{Furnas84}.
A full unrooted binary tree with exactly $n = 4$ leaves is also
referred to as \emph{simple quartet topology}, or just as
\emph{quartet}~\cite{Furnas84,graph2000}. Given a set $N$ of $n\geq 4$ objects, 
the number of sets of four objects from the set $N$ is given by:
${n \choose 4} = \frac{n!}{4!(n-4)!}=\frac{n(n-1)(n-2)(n-3)}{24}.$
Given four generic objects $\{a,b,c,d\} \in N$, there exist exactly
three different quartets: $ab|cd$, $ac|bd$, $ad|bc$, where the
vertical bar divides the two pairs of leaves, with each pair
labelled by the corresponding objects and attached to the same
internal node. 
Therefore the total number
of possible simple quartet topologies of $N$ is:
$3 \cdot {n \choose 4}.$
\begin{defn}
\textbf{- Consistency:} A full unrooted binary tree is said to be ``consistent'' with
respect to a simple quartet topology, say $ab|cd$, \emph{if and only
if} the path from $a$ to $b$ does not cross the path from $c$ to
$d$. This quartet $ab|cd$ is also said to be
``embedded'' in the given full unrooted binary tree.\end{defn}


Considering the set $N$, the MQTC problem accepts as input a \emph{distance matrix}, $\mat{D}$, which is a
matrix containing the dissimilarities, taken
pairwise, among the $n$ objects\footnote{It is therefore a symmetric $n
\times n$ matrix, with $n\geq 4$, containing non-negative reals, normalized between 0
and 1, as entries.}. 
To extract a hierarchy of clusters from the distance matrix, the
MQTC problem determines a full unrooted binary tree with $n$ leaves that visually represents the symmetric $n \times n$ distance matrix as well as possible 
according to a cost measure. 
Consider the set $Q$
of all possible  $3 \cdot {n \choose 4}$ quartets, 
and let $C: Q \rightarrow \Re ^+$ be a cost function assigning a real
valued cost $C(ab|cd)$ to each quartet topology $ab|cd \in Q$. The
cost assigned to each simple quartet topology is 
the sum of the dissimilarities (taken from $\mat{D}$) between each pair
of neighbouring leaves~\cite{CilibrasiNEW}. For example, the cost
of the quartet $ab|cd$ is
$C(ab|cd) = D_{(a,~b)} + D_{(c,~d)},$
where $D_{(a,~b)}$ and $D_{(c,~d)}$ indicate, respectively, the dissimilarities
among ($a$ and $b$) and ($c$ and
$d$), obtained from the $\mat{D}$.

Consider now the set $\Gamma$ of all full unrooted binary trees with
$2n-2$ nodes (i.e. $n$ leaves and $n-2$ internal nodes), obtained by
placing the $n$ objects to cluster as leaf nodes of the trees. 
For each $t \in \Gamma$, precisely one of
the three possible simple quartet topologies for any set of four
leaves is consistent~\cite{CilibrasiNEW}. 
Thus, 
there exist precisely ${n \choose 4}$ consistent quartet
topologies (one for each set of four objects) for each $t \in \Gamma$.

\begin{defn}
\textbf{- Cost function:} The cost associated with a full unrooted binary tree $t \in
\Gamma$ is the sum of the costs of its ${n \choose 4}$
consistent quartet topologies, that is:
$C(t)=\sum_{\forall ab|cd \in Q_t}{C(ab|cd)}$,
where $Q_t$ is the set of such ${n \choose 4}$ quartet topologies
embedded in $t$.\end{defn}

%

In a hierarchical clustering context, we do not even have a priori
knowledge that certain simple quartet topologies are objectively
true and must be embedded. Thus, the MQTC problem assigns a cost
value to each simple quartet topology, in order to express the
relative importance of the simple quartet topologies to be embedded
in the full unrooted binary tree having the $n$ objects as leaves.
The full unrooted binary tree 
with the minimum cost
balances the importance of embedding different quartet topologies
against others, leading to a binary tree that visually represents
the symmetric distance matrix $n \times n$ as well as possible. 
The solution of this problem allows the hierarchical representation of a
set of $n$ objects within a full unrooted binary
tree~\cite{Furnas84}. That is, the resulting binary
tree will have the $n$ objects assigned as leaves such that objects
with short relative dissimilarities will be placed close to each
other in the tree. This hierarchical clustering approach is also
referred in the literature to as \emph{quartet method}~\cite{CilibrasiNEW}. 
Here the MQTC problem was shown to be
NP-hard, 
and a Randomized
Hill Climbing heuristic was also implemented. 
Other MQTC metaheuristics based on Greedy Randomized Adaptive
Search Procedure, Simulated Annealing, and Variable Neighbourhood
Search were proposed 
in~\cite{myQuartet}.
In the following we discuss a possible 
exact solution approach containing some very interesting 
insights. It will be able not only to generate the first known benchmarks for the problem, but also 
to improve the current MQTC heuristics to date~\cite{CilibrasiNEW,myQuartet} by hybridization towards more efficient \textit{mat-heuristics}.

\section{An exact algorithm for the MQTC problem}
\label{sec-exact}


The approach can be dissected into 
two main tasks: \textbf{T1} - Given $n$ objects, generate all different full binary trees
topologies with $n$ leaves ($n- 2$ internal nodes)
\footnote{This sequence is referred in Sloane's Encyclopedia of Integer Sequences to as A000672.}; \textbf{T2} - For each different full binary tree topology, perform all
possible different permutations of its leaves, and find the
solution with the minimum cost for each considered topology.
The following are the fundamental blocks
of the proposed solution approach.

  - \textit{Definition of the initial topology structure}: Given $n$ objects and their pairwise dissimilarities stored in 
$\mat{D}$, the easiest possible initial topology structure
is the one formed by a single branch, i.e. where each $n-2$ internal nodes are connected with at least one leaf node.
Proceeding with the computations, our algorithm will increase the
complexity of the topology structure, 
discarding topologies already evaluated.

  - \textit{Data representation}:
To represent each full unrooted binary tree topology we make use of a Pseudo-adjacency matrix.
Given $n$ objects, each solution is represented by a $(2n-2) \times
(2n-2)$ matrix that we refer to as \emph{Complete Pseudo-Adjacency
matrix}, $\mat{A}=[A_{(i,~j)}]$ where $i,j=1,\ldots,(2n-2)$.
Similarly to an usual adjacency matrix (see e.g.~\cite{graph2000}),
in $\mat{A}$ we have $A_{(i,~j)} = 0$ for each non-diagonal entry
$A_{(i,~j)}$, $i \neq j$, where the two nodes $i$ and $j$ are not
directly connected each other. Conversely, when two nodes $i$ and
$j$ are directly connected we have $A_{(i,~j)} > 0$, but not the
value $1$ as in the usual adjacency matrix
representation~\cite{graph2000}. 
The Complete Pseudo-Adjacency matrix can be also partitioned as
\begin{align*}
\mat{A}^{(2n-2) \times (2n-2)}=\left[
\begin{array}{c:c}
    \mat{K}^{(n-2) \times (n-2)} & \mat{L}^{(n-2) \times n} \\ \hdashline
    \mat{L}'^{n \times (n-2)} & \mat{C}^{n \times n}
    \end{array}\right]
\end{align*}
where $\mat{K}$, referred to as \emph{Structure sub-matrix}, is a
$(n-2) \times (n-2)$ pseudo-adjacency matrix for the internal nodes
and it represents the structural topology of the given solution;
$\mat{L}$, referred to as \emph{Leaves sub-matrix}, is a $(n-2)
\times n$ adjacency matrix representing the connections of the
leaves with the internal nodes; $\mat{C}$, called \emph{Coefficients
sub-matrix}, is sized ${n \times n}$ and contains useful
coefficients to compute the cost of the solution.

- \textit{Computation of the cost function value}: Given a full unrooted binary tree $t$ with leaves assigned, to evaluate the cost of $t$ we should evaluate first all the ${n \choose 4} = \frac{n!}{4!(n-4)!}$ consistent quartet topologies embedded in its configuration, and then summing the costs of all these quartets, which are taken from the distance matrix $\mat{D}$. The cost of $t$ can be summarized as
$C(t) = \frac{1}{2} \sum_{ a,b=1 }^{ n}{  coef{(a,b)} \cdot D_{(
l_a,~l_b)} }$,
with $coef{(a,b)}$ denoting the coefficient multiplying the dissimilarity
${D( l_a,~l_b)}$. 
We can store the $coef{(\cdot,\cdot)}$ factors in the Coefficients
submatrix $\mat{C}$, such that, in this way, the cost of the full unrooted binary tree $t$ can be simply obtained in compact form by multiplying the transpose of the Coefficients submatrix $\mat{C}$ by the distance matrix $\mat{D}$, i.e.:
$C(t) = \frac{1}{2} \cdot \mat{C}^T \cdot \mat{D}$.

- \textit{Evaluation of all permutations of the leaves}:
Enumerating all permutations of the $n$ leaves attached to a specific full unrooted
binary tree topology takes a factorial time with respect to $n$. This factorial term is slightly soften by
avoiding the evaluation of the permutations of
two leaves attached to the same internal node, since this
permutation would not bring to any change in the cost value.
The factorial growth of this phase of the algorithm further confirms
the high computational complexity of the considered problem.
To some extent the adopted representation of data through the Complete
Pseudo-Adjacency matrix provides us some help. Indeed, to consider
all these combinations we can work directly on $\mat{A}$ and perform
all the possible permutations of rows and columns with
$i,j=(n-1),\ldots,(2n-2)$, i.e. rows and columns of the Leaves
sub-matrix, $\mat{L}$. The permutations will involve also commuting
the elements of the Coefficients sub-matrix, providing us
automatically the corrected coefficient values for the new
configuration, thus avoiding any additional calculation at this
purpose. After evaluated all the possible permutations, and
calculated the correspondent cost values, the configuration
providing the minimal overall cost will provide the best possible
solution obtained by the considered full unrooted binary tree.

- \textit{Generation of all different structural boron tree topologies}:
In order to enumerate and to generate all non-isomorphic full unrooted binary tree topologies with $n$ leaves, and to discard topologies already evaluated, we need to work only with the Structure sub-matrix $\mat{K}$, since we do not require any information on the particular connections order of the attached leaves.
Indeed $\mat{K}$ is a $(n-2) \times (n-2)$ pseudo-adjacency matrix for the internal nodes
and it represents itself the structural topology of the corresponding tree.
To discard topologies already evaluated, we make use of isomorphism invariants of graphs. That is, given two graphs $G_1$ and $G_2$, with adjacency matrices $A_1$ and $A_2$, these are isomorphic if and only if there exists a permutation matrix $P$ such that $P \cdot A_1 \cdot P^{-1}=A_2$. In particular $A_1$ and $A_2$ are similar and therefore have the same minimal polynomial, characteristic polynomial, spectrum (i.e., the set of its eigenvalues), determinant, and trace, which serve as isomorphic invariants of the graphs.
Following this concept, it is possible to discard incumbent full unrooted binary tree topologies if their isomorphic invariants coincide with those of the topologies already evaluated before. In this way useless repetitions are excluded.
At this point it is now possible to find the global optimum solution of the MQTC problem. For each
different full binary tree topology produced, we need to perform all possible
permutations of its leaves, as specified in the previous step, and then finding the solution with the
minimum cost for each considered topology. Finally, the minimal cost solution among this obtained set
will provide the global MQTC optimum.

\section{Summary and Outlook} \label{Conclusions}

We considered the minimum quartet tree cost (MQTC)
problem, a challenging novel graph optimization problem suited for
general hierarchical clustering. After providing the problem formulation, we scratched the details of an exact solution approach under current development\footnote{We are aware that a more detailed description of our algorithm is deemed necessary. 
We plan to include a deeper description of the method in an extended version of the paper, along with a thorough computational investigation.}.
Although the algorithm would able to get exact solutions only for relatively small
problem instances, due to the high complexity of this problem, it contains
some fascinating 
insights and novel concepts which may be of interest for the
improvement of the metaheuristics to date, or for the development of
new hybrid heuristics or matheuristics with higher performance. 



\begin{scriptsize}
\bibliographystyle{plain}

\end{scriptsize}

\end{document}